\documentclass[aps,prl,twocolumn,superscriptaddress,showpacs,reprint]{revtex4-1}
\usepackage{graphicx}
\usepackage{mathrsfs}
\usepackage{amsmath}
\usepackage{subfigure}
\usepackage{bm}
\usepackage{verbatim}
\usepackage{color}
\usepackage{xcolor}
\usepackage{hyperref}
\newcommand{\ket}[1]{|#1\rangle}
\newcommand{\nl}{n_{_\text{I}}}
\newcommand{\nll}{n_{_\text{II}}}
\newcommand{\nlll}{n_{_\text{III}}} 
\newcommand{\tz}{\tau_z}

\begin{document}
\title{
Unconventional Quantum Hall Effect and Tunable Spin Hall Effect in ${\rm MoS_2}$ Trilayers
}
\author{Xiao Li}
\affiliation{Department of Physics, The University of Texas at Austin, Austin, Texas 78712, USA}
\author{Fan Zhang}\email{E-mail: zhf@sas.upenn.edu}
\affiliation{Department of Physics and Astronomy, University of Pennsylvania, Philadelphia, PA 19104, USA}
\author{Qian Niu}
\affiliation{Department of Physics, The University of Texas at Austin, Austin, Texas 78712, USA}\affiliation{International Center for Quantum Materials, Peking University, Beijing 100871, China}

\date{\today}

\begin{abstract}
We analyze the Landau level (LL) structure and spin Hall effect in a ${\rm MoS_2}$ trilayer. Due to orbital asymmetry, the low-energy Dirac fermions become heavily massive and the LL energies grow linearly with $B$, rather than with $\sqrt{B}$. Spin-orbital couplings break spin and valley degenerate LL's into two time reversal invariant groups, with LL crossing effects present in the valence bands.
We find a field-dependent unconventional Hall plateau sequence 
$\nu=\cdots$ $-2M-6$, $-2M-4$, $-2M-2$, $-2M-1$, $\cdots$, $-5$, $-3$, $-1$, $0$, $2$, $4$ $\cdots$.
In a p-n junction, spin-resolved fractionally quantized conductance appears in two-terminal measurements with a controllable spin-polarized current that can be probed at the interface. We also show the tunability of zero-field spin Hall conductivity.
\end{abstract}
\maketitle

{\color{cyan}{\it{Introduction.}}}---
Successful isolation of a single molybdenum disulfide (${\rm MoS_2}$) trilayer presents a new platform  to explore interesting two dimensional (2D) electronic physics~\cite{Geim_NewPlatForm,Zhang_StrongPLinMLMOS2,Heinz_PLinMLMOS2,Kis_MOS2Transistor}.
The honeycomb lattice structure of a ${\rm MoS_2}$ trilayer  (when viewed from the top)  and its low-energy Dirac physics are reminiscent of graphene~\cite{Geim_GrapheneReview}. Indeed, the ${\rm MoS_2}$ trilayer exhibits advantages over graphene in several areas of intense recent interest, specifically, whether this 2D material has an energy gap~\cite{PNAS_NoGap,Allen_NoGap,Geim_NoGap} and whether it has substantial
spin-orbit couplings (SOC)~\cite{KaneMele_QSHinGraphene,YuguiYao_SOCgap,Allen_SOC}. Unfortunately, the answers for graphene to date are still not satisfactory, even though tremendous efforts have been made to improve the possibilities~\cite{Varykhalov_NiSubstrate,RuqianWu_doping}.

When the layered compound ${\rm MoS_2}$ is thinned down to a single trilayer, it departs from an indirect gap material to a direct gap material~\cite{GalliJ_DirectBandgapInMLMOS2,Eriksson_DirectBandgapInMLMOS2,Han_DFT,Heine_DFT,
Zhang_StrongPLinMLMOS2,Heinz_PLinMLMOS2}. Additionally, some density functional theory (DFT) calculations have shown that there exists large SOC in ${\rm MoS_2}$~\cite{Zhu_GiantSOCinMLMOS2,Xiao_TwobandModel,Eugene_SOCMOS2}.
The existence of a large energy gap and strong SOC may place this newly discovered 2D platform ahead of graphene in the race for the next generation of semiconductors. 
More recently a ${\rm MoS_2}$ transistor with room-temperature mobility about $200$\,cm$^2$/(V$\cdot$s) has already appeared~\cite{Kis_MOS2Transistor}.

In this Letter, for the first time, we analyze how SOC influences the Landau level (LL) spectrum of massive Dirac fermions and how to increase the spin Hall effect in a ${\rm MoS_2}$ trilayer.
We find that the Hall plateau is field-dependent and follows an unconventional sequence that has an even--odd--even transition.
The LL energies grow linearly with $B$, and the spin and valley degenerate LL's are broken into two time reversal invariant groups.
The broken symmetry in the valence band also gives rise to LL crossing effects, leading to pronounced peaks in the measurement of longitudinal magnetoresistance.
We further investigate the case of a p-n junction, where spin-resolved fractionally quantized conductance appears in two-terminal measurements, with a controllable spin-polarized current that can be probed at the interface by scanning tunneling microscope (STM). We also explicitly show how to tune and increase the zero-field spin Hall conductivity by reducing the inversion asymmetry.
None of these exotic features is able to be observed in graphene, even with the help of electron-electron interactions.

\begin{figure}[t]
\centering
\includegraphics[scale=0.49]{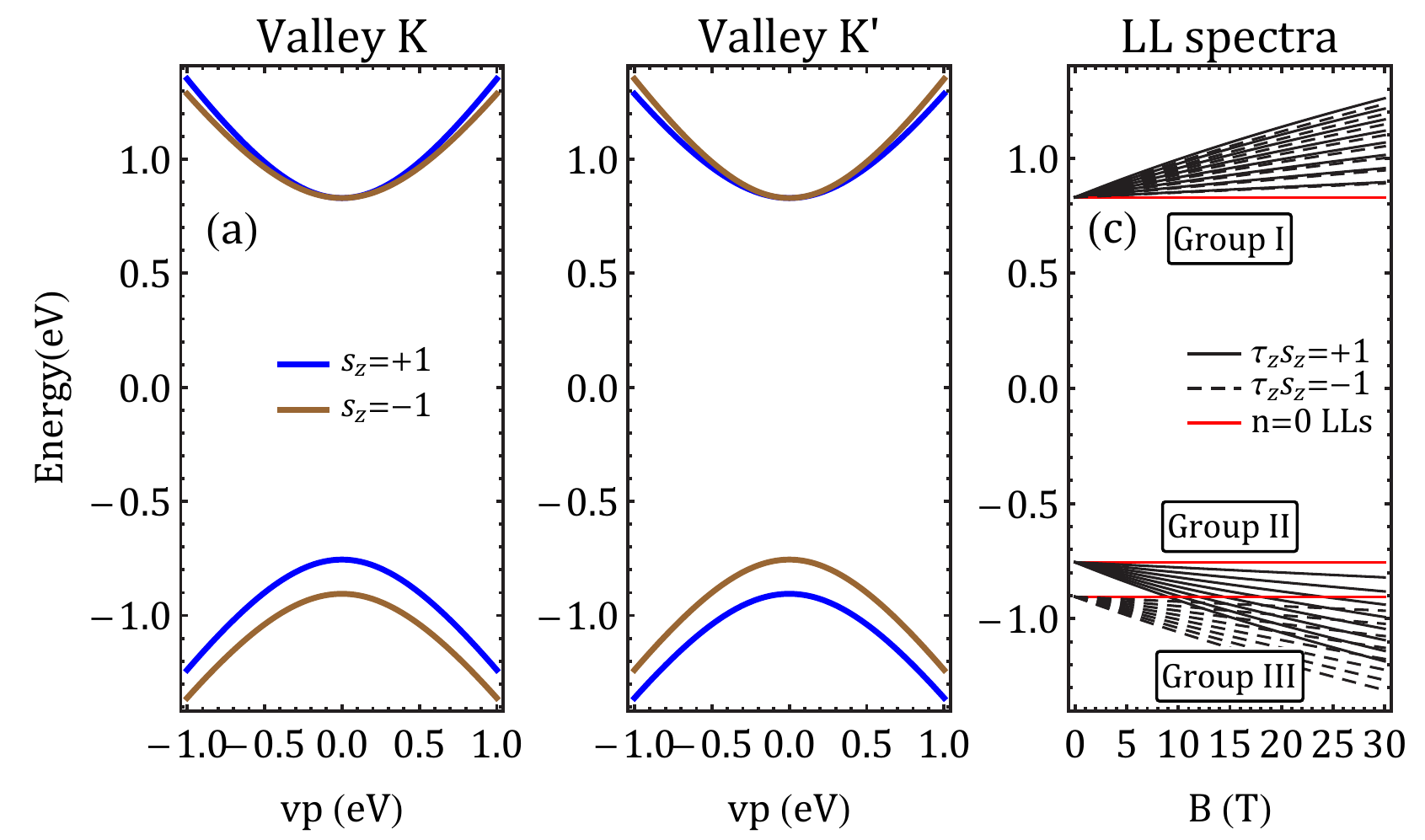}
\includegraphics[scale=0.39]{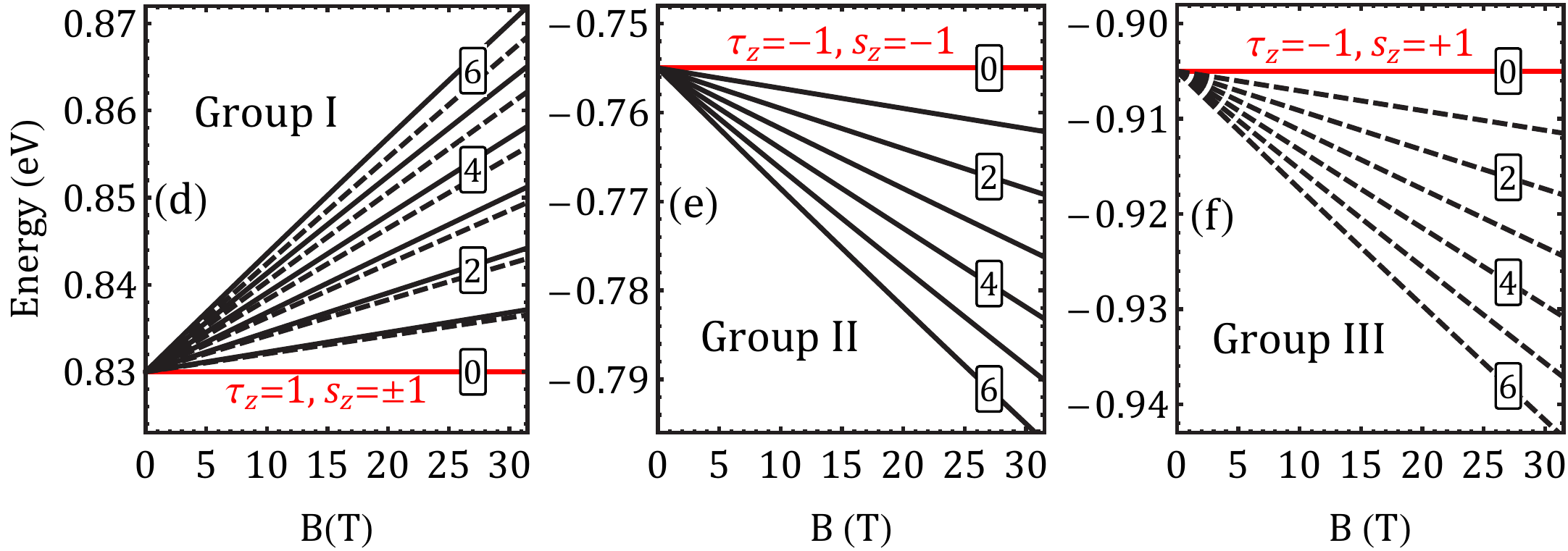}
\caption{\label{fig:Bandstructure}(color online). (a) and (b) Electronic band structure near valley K and K'. (c) LL's with $n=0,10,...,80$ orbitals. (d)-(f) Enlarged view of the LL's in Group I, II, and III in (c).
The $n\neq 0$ LL's are broken into $\tau_z s_z=1$ doublets and $\tau_z s_z=-1$ doublets, due to the SOC and inversion asymmetry. LL crossing occurs between group II and III.  The $\nl=0$ LL is spin degenerate and only appears at valley K. The $\nll,\nlll=0$ LL's are spin-filtered and appear only at valley K'.}
\end{figure}

{\color{cyan}{\it{Continuum theory.}}}---
We start from a description of the low-energy model of an isolated ${\rm MoS_2}$ trilayer, which applies generally to other group-VI dichalcogenides with the same crystal structure. The top and bottom ${\rm S}$ layers and the middle ${\rm Mo}$ layer are parallel triangular lattices. Because of their ABA relative stacking order, the top view of this trilayer forms a honeycomb lattice with ${\rm S}$ and ${\rm Mo}$ atoms at A and B sites, respectively. Near the Brillouin zone inequivalent corners K and K', the conduction and valence band states are approximately from $\ket{\phi_c} = \ket{d_{z^2}}$ and $\ket{\phi_{v}^{\tz}} =(\ket{d_{x^2-y^2}}+i\tz\ket{d_{xy}})/\sqrt{2}$ orbitals, respectively. This effective two-band model has been suggested by DFT calculations~\cite{Xiao_TwobandModel} and supported by optical experiments~\cite{Wang_Valleytronics,Xiao_SelectiveOccupationExp,Heinz_OpticalValleyPolarization,Sallen_Optical}.
To linear order in $p$, the effective ${k\!\cdot\! p}$ Hamiltonian in the above basis reads
\begin{align}\label{Eq:Hamiltonian}
\mathcal{H} = v(p_x\tau_z\sigma_x+p_y\sigma_y)+\Delta\sigma_z-\lambda\tau_z s_z\sigma_z+\lambda \tau_z s_z\,,
\end{align}
where the Pauli matrices $\bm\sigma$ operate on the space of the $d_{z^2}$ and $d\pm id$ orbitals, $\tau_z=\pm 1$ labels the K and K' valleys, and $s_z=\pm 1$ denotes the electron spin $\uparrow$ and $\downarrow$. The Fermi velocity $v$ is given by $v=at/\hbar=0.53\times10^6$\,m/s, where $t$ is the effective hopping between the two $d$-orbitals of ${\rm Mo}$ mediated by the $p$-orbitals of ${\rm S}$. As anticipated, the inversion asymmetry \cite{inversion} between $d_{z^2}$ and $d\pm id$ orbitals gives rise to the $\Delta\sigma_z$ mass term which pins the ground state to a quantum valley Hall (QVH) insulator~\cite{Xiao_graphene,Zhang_chiral_graphene}. ${\rm Mo}$ atoms provide strong intrinsic SOC $\sim\tau_z s_z\sigma_z$~\cite{KaneMele_QSHinGraphene,Zhang_chiral_graphene} that adjusts the energy gaps to $2(\Delta-\lambda)$ for $\tau_z s_z=1$ bands and to $2(\Delta+\lambda)$ for $\tau_z s_z=-1$ bands. Note that this SOC perturbation preserves inversion ($\mathcal{P}=\tau_x\sigma_x$) \cite{inversion} and time reversal ($\mathcal{T}=i\tau_x s_yK$) symmetries.
As a combined effect of breaking inversion symmetry and strong SOC, the $\mathcal{T}$ invariant term $\lambda \tau_z s_z$ breaks the particle-hole symmetry by oppositely shifting the $\tau_z s_z=\pm 1$ bands.  Using $\Delta=830$\,meV and $\lambda=37.5$\,meV extracted from DFT calculations~\cite{Xiao_TwobandModel}, Fig.~\ref{fig:Bandstructure} plots the band structure of a ${\rm MoS_2}$ trilayer which exhibits two features that substantially differ from graphene, i.e., (i) a large QVH band gap and (ii) a lifted degeneracy between $\tau_z s_z=\pm 1$ bands. We address that while the conduction band bottoms are lined up for all flavors, the valence band tops have a significant shift in energy between $\tau_z s_z=\pm 1$ bands. These symmetry breaking features in the spin-valley space can be further verified by the flavor-dependent energy dispersions
\begin{align}\label{Eq:KineticEnergy}
E_{\pm}=\lambda\tz s_z\pm \sqrt{v^2p^2+\left(\Delta-\lambda\tz s_z\right)^2}\,,
\end{align}
where $\pm$ stands for the conduction and valence bands.

{\color{cyan}{\it{Broken symmetry LL's and LL crossing effects.}}}---
In the presence of a uniform magnetic field perpendicular to a ${\rm MoS_2}$ trilayer, the 2D kinetic momentum $\bm{p}$ in Eq.~\eqref{Eq:Hamiltonian} is replaced by $\bm{\pi} =\bm{p}+e\bm{A}/c$. We now derive the LL spectrum using the Landau gauge $\bm{A}=(0, Bx)$. The operators $\pi=\pi_x+i\pi_y$ coincide with the lowering operators, satisfying $\pi\phi_n = -i(\hbar/\ell_B)\sqrt{2n}\phi_{n-1}$ and $\pi\phi_0=0$. Here $\ell_B=\sqrt{\hbar/(eB)} = 25.6/\sqrt{B[T]}$\,nm is the magnetic length and $\phi_n$ is the $n$th LL eigenstate of an ordinary 2DEG. This model is approximately valid when $\hbar v/\ell_B$ is smaller than the band width $\sim 500$~meV.
To focus on the influences from SOC and inversion asymmetry on the LL's, the relatively smaller effects including Zeeman couplings, disorders and Coulomb interactions are neglected. We obtain the flavor-dependent LL spectrum
\begin{align}\label{eq:LandauLevels}
E_{n,\pm}=\lambda\tau_z s_z\pm\sqrt{n\hbar^2\omega_c^2+\left(\Delta-\lambda\tau_z s_z  \right)^2}\,,	
\end{align}
where $\omega_c = \sqrt{2} v/\ell_B$ is the cyclotron frequency. The corresponding eigenstates with $n>0$ can be formally written as $(\phi_n, a_{n,s_z}^{\pm} \phi_{n-1})^T$ for valley K and $(b_{n,s_z}^{\pm}\phi_{n-1}, \phi_{n})^T$ for K'. For the anomalous $n=0$ LL's, the eigenstates are $(\phi_0,0)^T$ with energy $\Delta$ and $(0,\phi_0)^T$ with energy $-\Delta-2\lambda s_z$. This clearly shows that the $SU(4)$ invariant four anomalous $n=0$ LL's are broken into a two-fold spin degenerate conduction band $\nl=0$ LL at valley K and two spin split valence band $\nll=0$ and $\nlll=0$ LL's at K', as shown in Fig.~\ref{fig:Bandstructure}(c), leading to integer quantum Hall effects at $\nu=0$ and $\nu=-1$ but not $\nu=1$.
This is reminiscent to the anomalous $n=0$ LL's in few-layer graphene systems~\cite{Zhang_HundsRule}. In graphene the $SU(4)$ symmetry of $n=0$ LL's are completely lifted by electron-electron interactions~\cite{Ong_QHE,XuDu_QHE,Kim_QHE} while the particle-hole symmetry remains, whereas in ${\rm MoS_2}$ trilayer both the $SU(4)$ and particle-hole symmetries are broken by the SOC and the inversion asymmetry.

Other exotic features of LL's in ${\rm MoS_2}$ trilayers can be visualized in Fig.~\ref{fig:Bandstructure}(c)
and further understood by expanding Eq.~(\ref{eq:LandauLevels}) at $nB<50$\,T:
$E_{n,\alpha}=2\lambda\tau_z s_z\delta_{\alpha,-}+\alpha(\Delta+\frac{e\hbar v^2}{\Delta -\lambda\tau_z s_z})nB$ with $\alpha=\pm$.
(i) Because of the heavily massive Dirac Fermion character, the LL energy dependence on $B$ appears to be {\em linear} rather than $\sqrt{B}$. (ii) SOC break the LL's into two $\mathcal{T}$ invariant groups, with $\tau_z s_z=1$ and $\tau_z s_z=-1$. However, each $n\neq0$ LL is still doubly degenerate in each group, consisting of one spin $\uparrow$ state from one valley and one spin $\downarrow$ state from the other valley. (iii) The energies of two group LL's in the valence band not only have different slopes in $B$ but also shift rigidly at $B=0$, leading to LL crossing effects~\cite{Jarillo_QHEinTrilayerGraphene} at magnetic fields that satisfy
\begin{align}\label{eq:LLcrosing}
B_c = \dfrac{4\lambda(\lambda+\Delta)}{e\hbar v^2(\nll-\nlll)}
	+\dfrac{8\lambda^2 \nlll}{e\hbar v^2(\nll-\nlll)^2}\,,
\end{align}
where $\nll$ and $\nlll$ are the LL orbitals for Group II ($\tau_z s_z=1$) and Group III ($\tau_z s_z=-1$) shown in Fig.~\ref{fig:Bandstructure}(c). The second term is negligible when $\nll>2.7\nlll$.

\begin{figure}[t]
\centering
\includegraphics[scale=0.5]{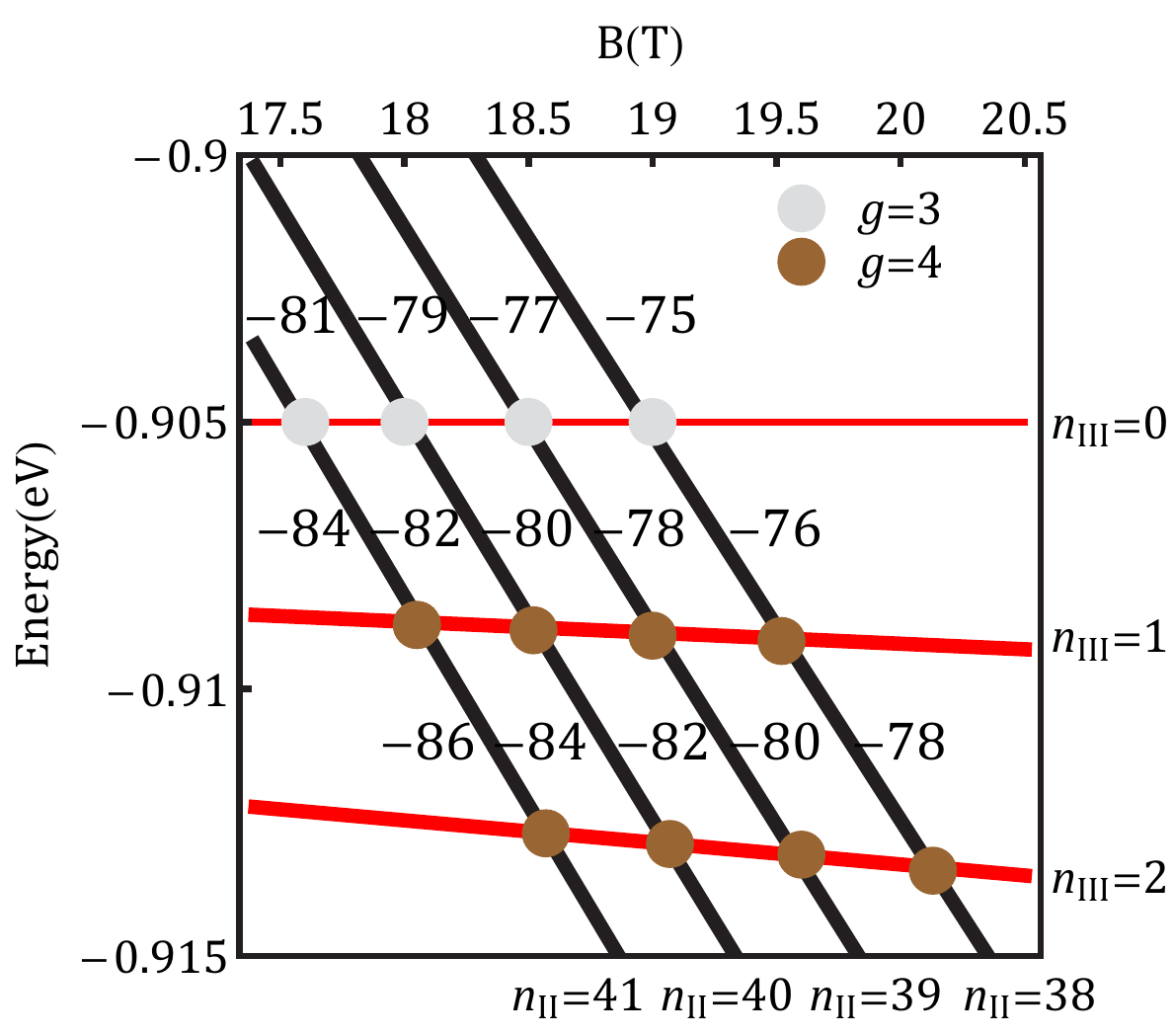}
\caption{(color online). Valence band LL crossing effects between LL's with $\nll=38, 39, 40, 41$ from group II (black) and LL's with $\nlll=0, 1, 2$ from group III (red). All LL's are doubly degenerate except that the $\nlll=0$ LL is non-degenerate. The negative number denotes the filling factor of the region bounded by the surrounding LL's. The crossing points have degeneracy $g=3$ at $\nlll=0$ and $g=4$ otherwise.\label{fig:LLcrossing}}
\end{figure}

As an example, in Fig.~\ref{fig:LLcrossing} we consider the crossings between LL's with $\nll=38, 39, 40, 41$ from group II and LL's with $\nlll=0, 1, 2$ from group III. The $\nlll=0$ LL is non-degenerate while other LL's are all doubly degenerate. As a consequence, each region bounded by three LL's above the $\nlll=0$ LL has an odd filling factor $\nu=-2(\nll+\nlll)-1$ while bounded by four LL's below has an even filling factor $\nu=-2(\nll+\nlll)$, where $\nll$ and $\nlll$ are the orbital indices of the right and the lower LL's. These filling factors are all negative because the crossings only occur in the valence band. In addition, the crossing points at the $\nlll=0$ LL all have degeneracy $g=3$, while other crossing points all have degeneracy $g=4$.
The crossing of two LL's results in increased degeneracies and lead to pronounced peaks in the measurement of longitudinal magnetoresistance, whose height is proportional to its degeneracy. Similar LL crossing effects in ABA-stacked trilayer graphene have been observed~\cite{Jarillo_QHEinTrilayerGraphene} via Shubnikov-de Haas oscillations.

Even in the absence of interactions, the unconventional Hall plateaus follow the integer sequence 
$\nu=\cdots$, $-2M-6$, $-2M-4$, $-2M-2$, $-2M-1$, $\cdots$, $-5$, $-3$, $-1$, $0$, $2$, $4$, $\cdots$,
where $M=\left[{4\lambda(\Delta+\lambda)}/{(eB\hbar v^2)}\right]$~\cite{IntegerM} reflects the fact that the $\nlll=0$ LL lies between the LL's with $\nll=M$ and $\nll=M+1$. The step of two in the sequence is a consequence of the $\tau_z s_z=\pm 1$ classification, a hallmark of SOC. A step-one jump reflects the filling of a $n=0$ LL. The presence of $\nu=0$ and $-1$ arises from inversion asymmetry that makes the Dirac fermions massive and from SOC that rigidly shifts the two groups of valence band LL's with different $\tau_z s_z$ characters. The switching between even and odd filling factors is a direct result of broken $SU(4)$ symmetry among the anomalous $n=0$ LL's. We address that the $\nll,\nlll=0$ LL's are {\em spin-filtered} and the exotic LL sequence is purely due to the non-interacting band structure.

\begin{figure}[b]
\centering
\includegraphics[scale=0.4]{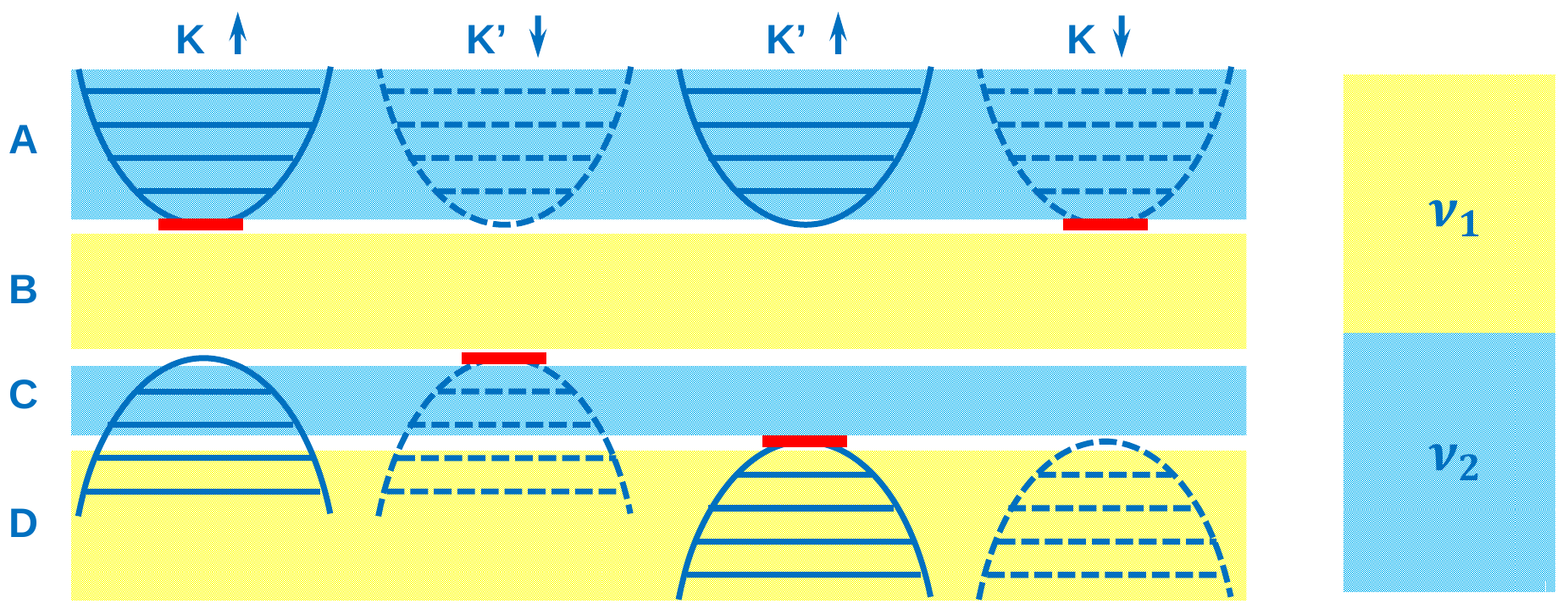}
\caption{\label{fig:PN} (color online). Left panel: The solid (dashed) curves represent spin $\uparrow$ ($\downarrow$) bands and the parallel lines denote their LL's. Regions A-D are energy windows separated by the three lifted energies of the four $n=0$ LL's that are depicted by red lines.
Right panel: a schematic p-n junction with two different filling factors $\nu_1$ and $\nu_2$ at two regions.}
\end{figure}
{\color{cyan}{\it{Spin-resolved p-n junctions.}}}---
A ${\rm MoS_2}$ p-n junction can be realized by using electrostatic gating to locally control the carrier type and density in two adjacent
regions. In such a device, transport measurements in the quantum Hall regime reveal new plateaus with integer and fractional filling factors of two-terminal conductance across the junction. This effect arises from the redistribution of quantum Hall current among spin-resolved edge channels propagating along and across the junction.
When $s_z$ is a good quantum number, because the edge channels of $\nll,\nlll=0$ LL's are spin-filtered, the full equilibrium must be achieved within each spin species separately. Consequently, the net conductance (in units of ${e^2}/{h}$) across the junction is quantized as follows
\begin{eqnarray}
G_{pp,nn}&=&\min\{|\nu_{1\uparrow}|,|\nu_{2\uparrow}|\}+\min\{|\nu_{1\downarrow}|,|\nu_{2\downarrow}|\}\,,\nonumber\\
G_{pn}&=&\dfrac{|\nu_{1\uparrow}|  |\nu_{2\uparrow}|}{|\nu_{1\uparrow}|+|\nu_{2\uparrow}|}
	+ \dfrac{|\nu_{1\downarrow}|  |\nu_{2\downarrow}|}{|\nu_{1\downarrow}|+|\nu_{2\downarrow}|}\,,
\end{eqnarray}
where $\nu_{1\uparrow}+\nu_{1\downarrow}=\nu_1$ and $\nu_{2\uparrow}+\nu_{2\downarrow}=\nu_2$. In this limit, for a junction with $\nu_1=2n$ and $\nu_2=-1$, electrons with one of the two spin-flavors are forbidden to flow into the p-doped region, since the only available LL $\nll=0$ is spin-filtered. Therefore, the net conductance is given by $G_{pn}=n/(n+1)$.
The conductance across the junction becomes spin-independent when $s_z$ is no longer a good quantum number, {\em e.g.}, due to strong magnetic disorders. This limit is similar to the case of graphene~\cite{Williams_GraphenePNJExp,Levitov_GraphenePNJTheory,Kim_GraphenePNPJunction} where all possible filling factors are even numbers because of the spin degeneracy.  Therefore, the corresponding conductance reads $G_{pn} =|\nu_1||\nu_2|/(|\nu_1|+|\nu_2|)$ in the bipolar regime or $G_{pp,nn}=\min\{|\nu_1|,|\nu_2|\}$ in the unipolar regime. Take the same example in which $\nu_1=2n$ and $\nu_2=-1$, the net conductance in this spinless limit becomes $G = 2n/(2n+1)$ instead.

In addition to the novel transport properties, STM probes at the interface of two junctions can also detect a special fingerprint of the spin-filtered $\nll=0$ LL, revealing the strongly broken symmetry between the two groups of LL's in the valence band.
As shown in Fig.~\ref{fig:PN}, when the Fermi energy of one region is fixed between the two valence band $n=0$ LL's (region C), namely, between $2\lambda-\Delta$ and $-2\lambda-\Delta$ indicated by Eq.~\eqref{eq:LandauLevels}, while the Fermi energy of the other is outside this energy window, there will be one spin-filtered chiral edge state, among all the $|\nu_1-\nu_2|$ channels, propagating along the interface. The chiral current will be {\em controllable} in the following senses. (i) Switching the magnetic field direction flips the spin-polarization of the current. (ii) Interchanging $\nu_1$ and $\nu_2$ switches the current direction while tuning $\nu_1$ and $\nu_2$ adjusts the current amplitude. (iii) Switching one of the Fermi level between A/B and D regions while fixing the other at C region changes the carrier type and flips the spin-polarization.

\begin{figure}[t]
\centering
\includegraphics[scale=0.45]{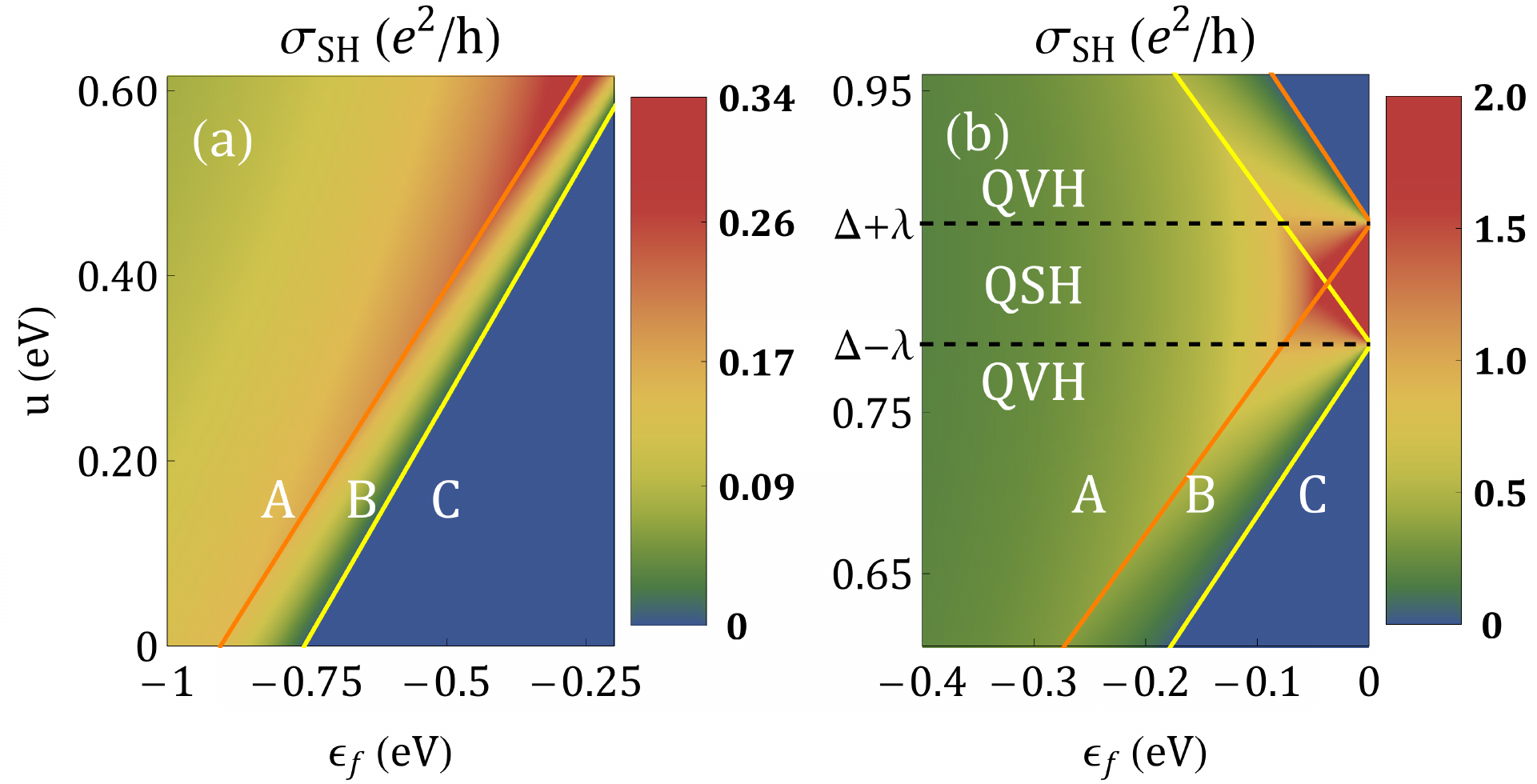}
\caption{\label{Fig:SpinConductivity} (color online). (a) Valence band spin Hall conductivity as a function of $\epsilon_f$ and $u$ that reduces inversion asymmetry. The yellow (orange) line denotes the top of the $\tau_z s_z=+1\,(-1)$ valence band. (b) Quantum phase transitions between QVH and QSH states at $u=\Delta\pm\lambda$.}
\end{figure}

{\color{cyan}{\it{Spin Hall conductivity.}}}---
Clearly shown in Fig.\ref{fig:PN}, the spin Hall conductivity is quantized to $\sigma_{SH}=e^2/h$ when the Fermi energy lies in the energy window C, due to the filling of spin-filtered $\nlll=0$ LL. We address that $\sigma_{SH}$ does not vanish even in the absence of fields~\cite{Sinova_UniversalIntrinsicSpinHall,Kato_SpinHallExp,Nagaosa_AHE,Shoucheng_SpinHall}, which is another consequence of the nontrivial valence band structure of ${\rm MoS_2}$ trilayers.
In a massive Dirac fermion model, the momentum space Berry curvature~\cite{Xiao-RMP,Zhang_chiral_graphene} in the valence band is nontrivial and reads
\begin{align}
\Omega_{\hat{z}}(\bm{k},\tau_z, s_z) =\dfrac{\tau_z\hbar^2 v^2 m}{2[v^2k^2+m^2]^{3/2}}\,,
\end{align}
where $m=\Delta-\lambda\tau_z s_z$ is the flavor-dependent mass. At zero temperature, we obtain the valence band spin Hall conductivity (in units of $e^2/h$) by integrating $s_z\Omega_{\hat{z}}(\bm{k},\tau_z, s_z)$ over the occupied states and summing over the spin-valley flavors.
As it happens in ${\rm MoS_2}$ trilayers, the inversion asymmetry dominates the SOC ($\Delta>\lambda$) and the system is pinned to a QVH insulator in which $\sigma_{SH}=0$ when the Fermi energy $\epsilon_f$ lies in the gap, implied by the $\tau_z$ dependence of $\Omega_{\hat{z}}$. Due to the mass difference between $\tau_z s_z=1$ and $\tau_z s_z=-1$ groups, however, $\sigma_{SH}$ contributions from the two groups are not completely canceled out when $\epsilon_f$ crosses valence bands. These features can be seen in the $u=0$ line trace in Fig.~\ref{Fig:SpinConductivity}(a). If in the opposite limit, assuming $\Delta=0$, the system becomes a quantized spin Hall (QSH) insulator with $\sigma_{SH}=2$ as shown by the middle zone in Fig.~\ref{Fig:SpinConductivity}(b). Therefore, if the inversion asymmetry could be compensated externally, $\sigma_{SH}$ will be enhanced as long as $\epsilon_f$ crosses the valence band, even though the system remains a QVH insulator. The inversion asymmetry arises from the difference between $d_{z^2}$ and $d\pm id$ orbitals, which can be possibly modulated by applying an electric field or chemical doping, straining or electric gating the two ${\rm S}$ layers. To investigate the tunability of $\sigma_{SH}$ by varying $\epsilon_f$ and by reducing inversion asymmetry,
we propose a linearized phenomenological theory, i.e., the momentum-independent terms in Eq.~(\ref{Eq:Hamiltonian}) is replaced by
\begin{align}\label{Eq:NewHamiltonian}
\mathcal{V}(\tau_z,s_z)=\lambda(1-\dfrac{u}{\Delta}) \tau_z s_z+(\Delta-u)\sigma_z-\lambda\tau_z s_z\sigma_z\,,
\end{align}
where $u$ is the modeled external potential that reduces the inversion asymmetry between $d_{z^2}$ and $d\pm id$ orbitals.
In the limit of $u=0$ this model recovers Eq.~(\ref{Eq:Hamiltonian}). In the special case of $u=\Delta$, inversion symmetry is restored and both $\sigma_z$ and $\tau_z s_z$ terms vanish.

Fig.~\ref{Fig:SpinConductivity} plots $\sigma_{SH}$ as a function of $\epsilon_f$ and $u$.
The yellow (orange) line maps out where $\epsilon_f$ touches the top of the valence band with $\tau_zs_z=+1$ ($\tau_zs_z=-1$). In region C where $\epsilon_f$ is inside the bulk gap, $\sigma_{SH}$ is identically $0$. When $\epsilon_f$ lies between the two valence band tops (region B), $\sigma_{SH}$ increases with decreasing $\epsilon_f$.
After $\epsilon_f$ drops far below the lower valence band top (region A), $\sigma_{SH}$ starts to decrease slowly. Overall, $\sigma_{SH}$ is maximized when $\epsilon_f$ is near the lower valence band top and grows substantially as $u$ increases. When $u$ is further increased, continuous quantum phase transitions (QVH-QSH-QVH) occur at $u=\Delta\mp\lambda$, as shown in Fig.~\ref{Fig:SpinConductivity}(b). Near the first (second) critical point, the mass of the Dirac fermions with $\tau_z s_z=1$ ($\tau_z s_z=-1$) changes sign and $\sigma_{SH}$ jumps from $0$ to $2$ ($2$ to $0$) when $\epsilon_f$ is inside the gap.

{\color{cyan}{\it{Discussions.}}}---
The Hall plateau in graphene follows the sequence $\nu=4(n+1/2)$, with $SU(4)$ symmetry breaking only in the case of high fields, weak disorders, and strong interactions \cite{Ong_QHE,XuDu_QHE,Kim_QHE}. In a ${\rm MoS_2}$ trilayer, quantum Hall ferromagnetism arises naturally even in the absence of interactions. SOC and inversion asymmetry also imprint a single-particle signature on the LL spectrum: the broken symmetry of the $n=0$ LL's and their energies are independent of the field strength. The unconventional Hall plateau sequence in a ${\rm MoS_2}$ trilayer becomes richer in a p-n junction with the appearance of
a spin-resolved fractionally quantized conductance. A controllable spin-polarized current materializes within this geometry and can
be probed by STM at the interface. Consequently, the MoS2 trilayer may find use
in a multi-functional spintronic device. Like in ABA trilayer graphene~\cite{Jarillo_QHEinTrilayerGraphene}, LL crossing effects occur in the ${\rm MoS_2}$ trilayer but only in its unconventional valence band. The two valleys are imbalanced in their contributions to $\sigma_{SH}$, providing hope for increasing $\sigma_{SH}$ by reducing the gap.
With improvements on its mobility, the ${\rm MoS_2}$ trilayer system may realize as an alternative
to graphene in fulfilling the desire for a gapped Dirac system with strong
SOC, and we also anticipate observations of its unique Hall phenomena discovered in this Letter.

{\color{cyan}{\emph{Acknowledgment.}}} ---
We are indebted to helpful discussions with R.~Hegde, W.~Bao, A.~H.~MacDonald, K.~F.~Mak, and Y.~Yao. X.~L. and Q.~N. are supported by DOE-DMSE (DE-FG03-02ER45958), NBRPC (2012CB-921300), NSFC (91121004), and the Welch Foundation (F-1255).
F.~Z. has been supported by DARPA under grant SPAWAR N66001-11-1-4110.



\begin{thebibliography}{99}
\bibitem{inversion}
 Here the orbital inversion symmetry is analogous to the parity inversion symmetry in graphene.
 For a ${\rm MoS_2}$ trilayer, this symmetry is only well defined in the low-energy model and is irrelevant in any lattice models.
\bibitem{IntegerM}
This sequence is general and shall be slightly modified at the LL crossing points. The analysis is similar to the example shown in Fig.~\ref{fig:LLcrossing}. The function $[x]$ here takes the integer part of the argument $x$.
\end{thebibliography}
\end{document}